\documentclass[aps,prb,twocolumn,showpacs]{revtex4}
\usepackage{graphicx}
\usepackage{epstopdf}

\begin{document}

\title{Magnetocaloric Study of Spin Relaxation in `Frozen' Dipolar Spin Ice Dy$_2$Ti$_2$O$_7$}
\author{M. Orend\'{a}\v{c}, J. Hanko, E.~\v{C}i\v{z}m\'{a}r, A.~Orend\'{a}\v{c}ov\'{a},}
 
\affiliation{Centre of Low Temperature Physics, Faculty of Science, P. J. \v{S}af\'{a}rik
University, Park Angelinum 9, 041 54 Ko\v{s}ice, Slovakia}
\email{orendac@kosice.upjs.sk}   
\author{ M. Shirai, S. T. Bramwell$^*$}
\affiliation{ Department of Chemistry, University College London, 20 Gordon Street, London WC1H 0AJ, United Kingdom\\
$^*$London Centre for Nanotechnology, 17-19 Gordon Street, London, WC1H OAH, United Kingdom}
\date{\today}

\begin{abstract}
The magnetocaloric effect of polycrystalline samples of pure and Y-doped dipolar spin ice 
Dy$_2$Ti$_2$O$_7$ 
was investigated at temperatures from nominally 0.3 K to 6 K and in magnetic fields of up to 2 T. 
As well as being of intrinsic interest, it is proposed that the magnetocaloric effect may be used as an appropriate tool for the qualitative study of slow relaxation processes in the spin ice regime. 
In the high temperature regime the temperature change on adiabatic demagnetization was found to be consistent with previously published entropy versus temperature curves. At low temperatures ($T < 0.4$ K) cooling by adiabatic demagnetization was followed by an irreversible rise in temperature that persisted after the removal of the applied field. The relaxation time derived from this temperature rise was found to increase rapidly down to 0.3 K. The data near to 0.3 K indicated a transition into a metastable state with much slower relaxation, supporting recent neutron scattering results. In addition, magnetic dilution of 50 \% concentration was found to significantly prolong the dynamical response in the milikelvin temperature range, in contrast with results reported for higher temperatures at which the spin correlations are suppressed. These observations are discussed in terms of defects and loop correlations in the spin ice state. 
\end{abstract}
\maketitle

\section{Introduction}
Geometrically frustrated magnets, in which the structure of the lattice leads to a competition between local spin - spin interactions, serve as model systems for the study of strongly correlated disorder~\cite{Ramirez}. Particular interest has been devoted to the spin ice materials Dy$_2$Ti$_2$O$_7$, Ho$_2$Ti$_2$O$_7$ and Ho$_2$Sn$_2$O$_7$, on account of the analogy between the orientation of their magnetic moments and the arrangement of O-H bonds in common hexagonal water ice~\cite{Harris1,Ramirez1,Kadowaki, Pauling}. In these materials, the magnetic rare earth ions, Dy$^{3+}$ or Ho$^{3+}$, reside on a lattice of corner-sharing tetrahedra. The crystal electric field introduces an almost perfect Ising anisotropy with the anisotropy axis oriented along the local $[111]$ crystallographic axis, a member of the cubic  $\langle 111\rangle$ set,  that points from a vertex of the tetrahedron to its centre. The nearest-neighbour spin-spin interaction is ferromagnetic and predominantly of dipolar origin: it leads to a strong frustration
with local spin configurations obeying ``ice rules'' - two spins oriented outward and two spins oriented inward, on each tetrahedron. As in the case of water ice, these rules stabilise a macroscopically degenerate ground state and prevent the system from long-range ordering. Indeed, no magnetic order is observed in the spin ice materials at least down to a temperature of 50 mK~\cite{Harris1,Fukazawa} and the observed residual entropy is in excellent agreement with the expected zero-point entropy of water ice, $1/2R\ln(3/2)$~ \cite{Ramirez1}. Various novel magnetic states in spin ice have been the subject of extensive theoretical \cite{Siddhartan1,Zhitomirsky,Moessner1,Ruff,Isakov} and experimental \cite{Sakakibara1,Saito,Fennell,Higashinaka} studies. 

The question of how the spin ice state is formed has stimulated several investigations of the magnetic dynamics of the spin ice materials with particular emphasis on relaxation phenomena~\cite{Ehlers,Matsuhira2,Matsuhira,Snyder2}. The magnetic relaxation in Dy$_2$Ti$_2$O$_7$ has been discussed in most detail. Ac-susceptibility studies of powdered Dy$_2$Ti$_2$O$_7$ up to 20 K suggest rather a wide distribution of relaxation times~\cite{Matsuhira}, but other studies \cite{Snyder2} suggest that spin ice responds to external magnetic field in a very limited range of relaxation times, an effect that has been ascribed to the high degree of chemical and structural order. In addition, ac-susceptibility data for Dy$_2$Ti$_2$O$_7$ indicate strongly frequency dependent spin freezing at about 16 K. This  was initially discussed in terms of collective dynamics, \cite{Snyder2,Snyder3}, but subsequently, on the basis of systematic studies of Dy$_{2-x}$Y$_x$Ti$_2$ O$_7$ in wide range of dilutions $x$~\cite{Snyder4} as well as neutron scattering~\cite{Steve1}, ascribed to a single spin flip process. 

Relaxation processes in the temperature range where the ice-like spin - spin correlations are already developed ($T < T_{\rm ice} \approx$ 2 K) have been addressed in much less detail. A recent neutron scattering investigation of Dy$_2$Ti$_2$O$_7$ and Ho$_2$Ti$_2$O$_7$ in an applied magnetic field \cite{Steve2} suggested the existence of metastable states and very slow dynamics governed by magnetic field. The effect of magnetic dilution for $T < T_{\rm ice}$ also remains unclear. In the present work we study the relaxation processes in both pure and diluted spin ice in a temperature range  where single spin flips must be significantly restricted by the well-established ice rules. We propose that,  due to the expected (slow) time  scale of the relaxation, the study of the magnetocaloric effect may serve as an alternative technique for the qualitative estimation of the temperature and magnetic field dependence of relaxation times. 
 
Magnetocaloric effects in strongly frustrated magnetic systems are, furthermore, of intrinsic interest, as the large degeneracies of these systems lead to special properties that are 
absent in nonfrustrated magnets. Specifically, it was suggested that if residual entropy is suppressed by magnetic field, then, during adiabatic demagnetization, the system will regain the entropy and an enhanced cooling effect may occur~\cite{Zhitomirsky1}. 
A large magnetocaloric effect in frustrated magnets has also been predicted to occur near the saturation field due to a macroscopic number of local modes, which, below the saturation field, remain gapless  \cite{Zhitomirsky2}. An enhanced cooling rate was observed in the Heisenberg pyrochlore  Gd$_2$Ti$_2$O$_7$ \cite{Sosin}, in agreement with theoretical predictions; the most efficient cooling appearing near the crossover between saturated and spin-liquid phases. For Dy$_2$Ti$_2$O$_7$, the 
magnetic field variation of the entropy was examined using the magnetocaloric effect in a magnetic field oriented in the $[111]$ direction, in order to characterise the ``giant entropy spike'' that accompanies
the  ice-rules breaking spin-flop transition~\cite{Aoki}. The magnetocaloric effect of Dy$_2$Ti$_2$O$_7$ associated with demagnetization to the zero field spin ice state has not been reported in detail before \cite{Flood}, but as discussed further below, poses an interesting and fundamental problem for thermodynamics.

In the present work we apply adiabatic demagnetization to study the response  of pure and magnetically diluted spin ice in slowly changing magnetic field at temperatures $T<T_{\rm ice}$. In both systems a recovery towards a new equilibrium state after finishing the demagnetization is clearly manifested. The observed response of Dy$_2$Ti$_2$O$_7$ enables the estimation of the temperature dependence of the relaxation time. The time dependence of the thermal response of (DyY)Ti$_2$O$_7$ suggests a more complicated relaxation process. It is found that magnetic dilution of    50 \% concentration significantly prolongs the relaxation time, in contrast with corresponding results obtained for $T>T_{\rm ice}$~\cite{Snyder1}.

\section{Experimental}

Polycrystalline powdered samples of Dy$_2$Ti$_2$O$_7$ and (DyY)Ti$_2$O$_7$  were prepared using well established solid-state synthesis techniques~\cite{from_Steve}. The Y$^{3+}$ ion was used for magnetic dilution since it has nearly the same ionic radius as Dy$^{3+}$. Both samples were confirmed, using X-ray diffraction, to be monophasic with the cubic pyrochlore structure. The magnetocaloric effect was investigated using two experimental devices. Adiabatic demagnetization in the milikelvin temperature range was conducted in a commercial dilution refrigerator TLE 200, whereas a home made $^{4}$He cryostat was applied for demagnetization above 1 K. Both Dy$_2$Ti$_2$O$_7$ and (DyY)Ti$_2$O$_7$  were in the form of coin shaped pellets of weights 495 mg and 502 mg, respectively. In order to minimise demagnetizing effects, the magnetic field was aligned with the  
flat aspect of the pellet.  The samples were connected to a metal frame using nylon threads of diameter 0.08 mm and length 2.5 cm. During the demagnetization the temperature of the frame followed 
the actual temperature of the samples to suppress the heat transport via threads. The temperature of the samples was monitored using 1 k$\Omega$  RuO$_{2}$ thermometers of RC550 type, mounted on the sample. These thermometers are characterized by low magnetoresistance \cite {Mark,Uhlig} and specific heat \cite{Cryogenics}. Considering the magnetoresistance of the thermometers used and the fact that a decrease of the temperature was induced by decreasing magnetic field, the magnetoresistance effects could be safely neglected in the temperature and magnetic field range of presented experimental data.

\section{Results}
In the first step adiabatic demagnetization was performed from 6 K to 0.7 K using various initial magnetic fields up to 2 T. In this temperature and magnetic field range ac-susceptibility studies \cite{Snyder3} suggest a relaxation time smaller than nominally 1 s. The adiabatic change of temperature with changing magnetic field is governed by thermodynamic relation
\begin{equation}
\left( \frac{\partial T}{\partial B} \right )_S = - \frac {T} {C} \left ( \frac{\partial S} { \partial B} \right )_T 
\end{equation}
where $C$ represents specific heat in field $B$ and the quantity $ ({\partial T} / {\partial B})_S$  can be obtained from the demagnetization. The relation (1) enables an indirect verification of the change of entropy with temperature and magnetic field, a relationship that can alternatively be determined from specific heat measurements.  Although direct determination of the values of the entropy from a magnetocaloric study is not possible, the adiabatic change of temperature with sweeping magnetic field can be predicted from the temperature and magnetic field dependence of the entropy and then compared with corresponding experimental data. This was done as follows.
During the demagnetization the temperature of Dy$_2$Ti$_2$O$_7$ was monitored as a function of swept magnetic field. On the obtained dependence, specific points $[T_i , B_i]$ were selected for those values $B_i$, for which the temperature dependence of the entropy $S(B_i,T)$ (see Fig. 1) was derived from specific heat data~\cite{Ramirez1}. The entropy values $S(B_i, T_i)$ were then read off  the reported entropy versus temperature curve for the given field $B_i$.  Considering that adiabatic demagnetization represents an isentropic process, the values of magnetic entropy obtained in each run were averaged. The comparison of the entropy estimated from specific heat data \cite{Ramirez1} and that obtained from demagnetization is presented in Fig. 1. The obtained reasonable agreement may serve as indirect confirmation of temperature and magnetic field dependence of entropy  proposed  from calorimetric data for Dy$_2$Ti$_2$O$_7$.
   
The release of residual entropy by magnetic field removal in geometrically frustrated magnets creates a rather unusual situation for magnetocalorimetry. Specifically, the formal adoption of Fig. 1 below 2 K and 1 T enables one to consider a state in appropriate magnetic field for which the entropy is much smaller than that in zero field (Fig. 2). Indeed, according to Fig. 2, such a state can be reached by adiabatic magnetization from the initial state (1) to an intermediate state (2) and subsequent cooling in constant magnetic field to the final state (3). Unlike in the case of a conventional magnet, the minimum temperature obtainable by demagnetization from state (3) can not be predicted from the temperature dependence of the entropy in zero magnetic field. This fact motivated subsequent magnetocaloric studies in the milikelvin temperature range. 

The adiabatic demagnetization below 1 K was studied using different sweep rates varying from 1.5 mT/min to 0.22 T/min with initial temperature 0.84 K and magnetic field 0.75 T. The selected initial temperature is higher than the value at which the bifurcation of field-cooled and zero field cooled magnetization data was observed~\cite{Snyder5}. In addition, the selected initial magnetic field is smaller than that which releases the whole residual entropy \cite{Ramirez1}. 
The time dependence of the temperature during demagnetization of Dy$_2$Ti$_2$O$_7$ and (DyY)Ti$_2$O$_7$  is presented in Fig. 3 and Fig. 4, respectively. The thermal response of Dy$_2$Ti$_2$O$_7$ is characterized by a linear decrease of temperature with a change of slope occurring in a relatively small time range. A small upturn of the temperature is observed towards the end of each run, in finite field. In contrast, (DyY)Ti$_2$O$_7$  is cooled with gradually increasing speed to a minimum temperature higher than that obtained for Dy$_2$Ti$_2$O$_7$ and also with a more significant temperature upturn. It should be stressed that in both materials the temperature increase continues even after finishing the demagnetization (see Inset in Fig. 3 and Fig. 4). Whereas for Dy$_2$Ti$_2$O$_7$ the temperature increase after the demagnetization seems to be of exponential form (see Fig. 3 (inset)), a more complicated dependence was observed in (DyY)Ti$_2$O$_7$ . 

The dependence of temperature on magnetic field during demagnetization for both materials and various sweep rates is presented in Fig. 5a. and Fig. 5b. The thermal response of Dy$_2$Ti$_2$O$_7$ on change of magnetic field is not very sensitive to sweep rate, at least for the used range of sweep rates:  only below 0.4 K is a mild dependence on sweep rate observed. In addition, for all used sweep rates, the change of slope in the temperature decrease occurs in a small range of magnetic fields around 0.3 T.  In contrast, a pronounced effect of sweep rate is demonstrated in (DyY)Ti$_2$O$_7$ , where the sweep rate significantly influences not only values of temperature during cooling but also the response after the magnetic field reaches zero value. The simplest explanation of this ``memory'' of the applied sweep rate is that the evolution of the system after removal of the field is limited by the temperature (and hence phonon population) of the lattice, as discussed further below. 
   
\section{Discussion}
The data presented here enable one to study the relaxation process in spin ice at temperatures lower than those for which ac-susceptibility data are available \cite{Snyder1,Snyder3} and to compare the influence of magnetic dilution on spin relaxation in the milikelvin temperature range with the detailed investigations reported for $T > T_{\rm ice}$~\cite{Snyder4}. The thermal response of Dy$_2$Ti$_2$O$_7$ on decreasing magnetic field suggests that above $\sim0.4$ K the magnetic and lattice/thermometer subsystems are in thermal equilibrium during the demagnetization. Below this temperature it is clear that at least part of the magnetic subsystem falls out of equilibrium with the lattice and remains at an effectively higher temperature. The final rise in temperature represents the equilibration of the whole system. The fact that the temperature continues to rise after demagnetization identifies this as a spontaneous and irreversible thermodynamic process that increases entropy. These changes are represented schematically by processes $3 \rightarrow 4$ and $4 \rightarrow 1$ on Fig. 2. We now discuss briefly several details of this process.
    
The change of slope in cooling rate observed near 0.3 T regardless of sweep rate could be associated with the formation of the kagome-ice state in the grains for which the magnetic field is oriented close to a $\langle 111\rangle$ direction. Specifically, magnetization studies of Dy$_2$Ti$_2$O$_7$ with field oriented along $[111]$~ \cite{Sakakibara1,Matsuhira3} revealed that the system is first magnetized into this intermediate state, still governed by ice rules, and where the degeneracy remains in the kagome plains oriented perpendicular to $[111]$. A sufficiently high magnetic field breaks the ice rules and drives the system into the ordered state. The resulting magnetization curve displays a plateau in the range 0.3 T $< B < $ 0.9 T, \cite{Sakakibara1} which also leads to a plateau in the field dependence of the residual entropy observed from nominally 0.3 T to 0.7 T~\cite{Matsuhira3}. Considering the random orientation of grains in the pellet,  it is conceivable that, for selected grains, the orientation of magnetic field will be close to a $\langle 111\rangle$ direction. These grains will not contribute to cooling during demagnetization with the aforementioned initial condition due to the presence of the plateau in the residual entropy.  Only when magnetic field becomes smaller than 0.3 T, will Dy$_2$Ti$_2$O$_7$ in all grains be driven back into spin ice state and so cooling becomes more efficient.

The time and magnetic field dependence of temperature below 0.4 K in Dy$_2$Ti$_2$O$_7$ suggests that the system becomes out of equilibrium on the time scale of measurement. It is apparent that, after finishing the demagnetization, the lattice of Dy$_2$Ti$_2$O$_7$ still accepts energy from the magnetic subsystem until thermal equilibrium is re-established at a certain temperature. This resulting temperature $T_{\rm final}$ is considered as a temperature reached by the demagnetization for a given sweep rate. The dependence of $T_{\rm final}$ on the sweep rate is presented in Fig. 6. 

As intuitively expected, slower demagnetization enables Dy$_2$Ti$_2$O$_7$ to be cooled to lower temperatures. However, when the resulting temperature approaches 0.3 K, its value does not seem to be so sensitive on the relative change of the sweep rate as observed for higher temperatures. The relaxation time $\tau$ for a given sweep rate can be estimated by fitting the thermal response after finishing the demagnetization using single exponential in the form :
\begin{equation}
 T(t) = T_0 + \Delta T \left( 1-exp \left(-\frac {t} {\tau} \right) \right)
\end{equation}
where $T_0$ represents temperature of the sample when magnetic field reaches zero value and $\Delta T $ denotes the temperature increase during the relaxation. 
The results of this analysis of the thermal response  are presented in Fig. 7. 
In the analysis, the approximation used represents the main source of error in the estimated values of the relaxation time. 
 
Whereas for sweep rates higher than 50 mT/min the estimated relaxation time seems to be about tens of seconds, a very pronounced increase occurs for lower sweep rates. Both the behaviour of the resulting temperature $T_{\rm final}$ and the sharp increase of relaxation time for the lowest sweep rates indicate a pronounced slowing down in relaxation when the system is demagnetized towards 0.3 K. This is manifested in the dependence of the relaxation time on the resulting temperature, see Fig. 8, which confirms a steep thermal dependence of the relaxation in the studied temperature range. We may compare this behaviour with that of a paramagnetic Kramers ion, which typically may relax by the direct, Raman and Orbach processes~\cite{Orbach,ScottJeffries}. 
As shown on the figure, the temperature dependence is close to the $T^{-9}$ dependence expected for the Raman process. In a magnetic field the direct process is also likely (it increases at $B^4$), but we expect the single ion Orbach process to be extinct in this temperature range~\cite{OrbachComment}. As the typical phonon energy is $\sim kT$ and the energy of a creation of an ice rules defect is $\sim kT_{\rm ice}$, the fact that cooling by phonon energy loss persists to $T \ll T_{\rm ice}$ suggests a low energy channel of spin relaxation: for example diffusion of ice rules defects or collective ``looped'' spin reversals~\cite{Melko}. An alternative mechanism that one might consider is a low energy collective spin reversal mediated by the transient generation of a defect in the spin ice
state. Such a mechanism would be Orbach-like, relying on phonons in a narrow band of energies around $\Delta \sim kT_{\rm ice}$ and so would have an Arrhenius temperature dependence with that activation energy. The Arrhenius form indeed gives a satisfactory description of the data with an activation energy of this magnitude (see Fig. 8). More theory and experiment is clearly required in order to distinguish between these candidate spin-lattice relaxation processes. 

The results of Fig. 8 may be compared with those obtained by ac-susceptibility~\cite{Snyder1}. However, quantitative comparison should be accepted with caution since, unlike in ac-susceptibility measurement, the relaxation time was estimated from a small, but non-negligible rise of the temperature.    
As illustrated in inset of Fig. 8  the presented results may be considered as a qualitative extension of ac-susceptibility studies towards 0.3 K.

At temperatures close to 0.3 K, there appears to be a more rapid increase of relaxation time (see Fig. 8), suggesting a crossover into a metastable state with much slower relaxation. Obviously a more convincing conclusion could be drawn if more data below 0.3 K were available. However, such a study should be performed in a significantly larger time scale that would require a new design of experimental setup: consequently it will be  subject of a future effort. The suggested crossover is nevertheless consistent with recent neutron scattering results for Dy$_2$Ti$_2$O$_7$, \cite{Steve2} where the existence of two regimes was proposed. Both regimes are characterized by spin correlations governed by ice rules, but whereas at higher temperatures ($T > ~$0.3 K) dynamical processes are still present, for lower temperatures ($T < ~$0.3 K) the system enters a frozen state in which the probability of spin flips becomes much smaller. A recent theoretical analysis of the zero field neutron scattering results~\cite{Tom}  also suggests why the final irreversible entropy increase ($4 \rightarrow 1$ on Fig. 2) is exothermic ($\Delta U < 0$).  In particular, the equilibrium state at 0.3 K appears to consist of only closed loops of spins that carry zero magnetization and are stabilised by further neighbour exchange coupling. These interactions combine with the dipolar interaction to give an energy scale of $\sim 0.3$ K that  possibly provides the necessary $\Delta U$. 

We finally consider the effect of magnetic dilution on the relaxation below
$T_{\rm ice}$. Although in (DyY)Ti$_2$O$_7$ a more complicated thermal response
after the demagnetization does not enable an analysis similar to that performed
for Dy$_2$Ti$_2$O$_7$, the observed sensitivity of the demagnetization of
(DyY)Ti$_2$O$_7$ on sweep rate immediately suggests an increase of relaxation
time with 50 \% magnetic dilution. This behavior contrasts with the observation
for $T > T_{\rm ice}$, where a non-trivial dependence of relaxation time as a
function of dilution was observed. Specifically, a concentration of magnetic
dilution smaller than nominally 12 \% was found to accelerate the dynamical
response~\cite {Snyder4}. That decrease of the relaxation time was ascribed to a
reduction of the local fields around each Dy$^{3+}$ ion by nonmagnetic dilution
and a subsequent smaller energy splitting of different spin states which makes
quantum tunneling of individual spins more probable. In contrast, for
concentrations of Y$^{3+}$ higher than 12 \%, an increase of the relaxation
time with increasing level of doping was observed. That phenomenon was ascribed
to an increasing energy barrier due to modifications in crystal field splitting
stimulated by dilution-induced changes of lattice parameters. The competing
effect of both mechanisms leads to the situation that for $T > T_{\rm ice}$ and
zero magnetic field the relaxation times of Dy$_2$Ti$_2$O$_7$ and
(DyY)Ti$_2$O$_7$ are nearly the same.
Although for (DyY)Ti$_2$O$_7$ the latter mechanism might also be present in the
low temperature regime, the observed behaviour seems to be more similar to that
of the 'collective paramagnet' Tb$_2$Ti$_2$O$_7$~\cite{Gardner}, for which spin
freezing around defects has been proposed. The pronounced difference in the
dynamical response of diluted spin ice may be understood either by the pinning
of diffusing defects (the absence of a spin creates an energy trap) or by the
pinning of spin loops. It may also be that the effectiveness of the phonon
population in maintaining equilibrium at low temperature is adversely affected
by the introduction of quenched chemical disorder. The systematic study of spin
relaxation in spin ice at $T < T_{\rm ice}$ with different concentrations of
nonmagnetic ions is clearly desirable in order to clarify the proposed
scenario.

\section{Conclusion}
The study of the magnetocaloric effect in Dy$_2$Ti$_2$O$_7$ above 2 K has
enabled verification of the temperature and magnetic field dependence of the
entropy calculated from experimental specific heat data. Adiabatic
demagnetization of spin ice in the  milikelvin temperature range has
demonstrated that the technique can be used to study slow relaxation processes.
The demagnetization of Dy$_2$Ti$_2$O$_7$ strongly suggests that down to 0.3 K
the dynamics are dominated by thermally activated spin-lattice relaxation in
which the characteristic relaxation time increases  with decreasing temperature
in a manner consistent with either a Raman process or an Orbach-like
process with activation energy $\sim kT_{\rm ice}$. In addition, at $T \approx
~$0.3 K the transition to a regime with much slower dynamics was indicated,
supporting the conclusion of neutron scattering experiments. It was also found
that for $T < T_{\rm ice}$, magnetic dilution of 50 \% concentration
significantly increases the relaxation time in contrast with reported behavior
in 'molten' spin ice. Future experimental studies, including adiabatic
demagnetization of single crystals, neutron scattering under adiabatic
conditions, studies at lower temperature and systematic studies of dilution,
will be necessary to clarify the various microscopic processes suggested by
these experimental results.

\begin{acknowledgments}
We would like to thank R. Moessner, T. Honecker and P. C. W. Holdsworth for enlightening discussions. In the early stage of this work we benefited from the interactions with M. W. Meisel and P. Koll\' ar. Expert technical help of M. Kaj\v nakov\'{a} is gratefully acknowledged. The work was supported by project VEGA 1/3027/06 and by ESF research network program HFM8.
Material support of U.S. Steel is also gratefully acknowledged.
\end{acknowledgments}

\begin{figure} [p]
\includegraphics{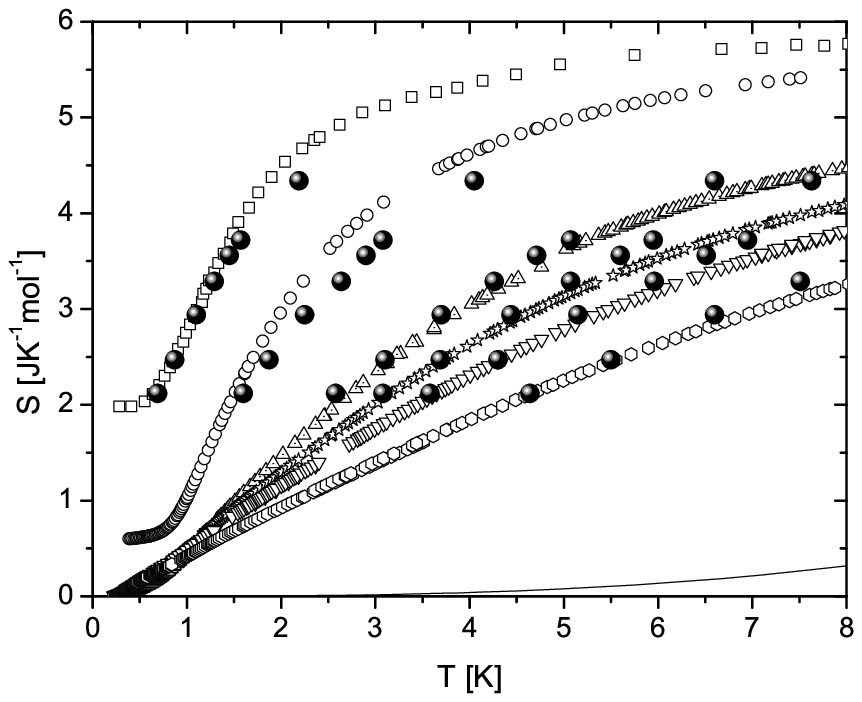}  
\caption{Temperature dependence of the entropy of powdered Dy$_2$Ti$_2$O$_7$. Open symbols denote entropy calculated from magnetic specific heat \cite{Ramirez1} in various magnetic fields (squares $B$=0 T, circles $B$=0.5 T, triangles up $B$=1 T, stars $B$=1.25 T, triangles down $B$=1.5 T, hexagons $B$=2 T). Full circles represents data obtained from adiabatic demagnetization. The full line denotes the entropy estimated from lattice specific heat.}
\label{fig. 1}
\end{figure}

\begin{figure}
\includegraphics{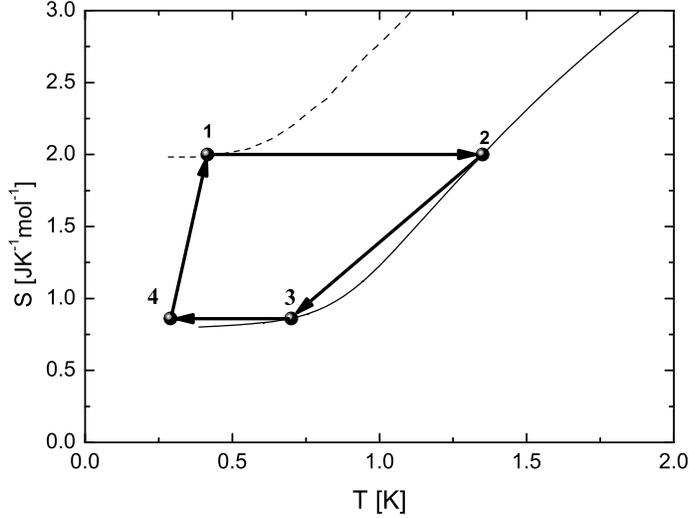} 
\caption{Temperature dependence of entropy of Dy$_2$Ti$_2$O$_7$ below $T$ = 2 K in $B$ = 0 (dashed line) and B = 0.5 T (solid line). Numbered full circles denote an initial state (1), an intermediate state (2) reached by adiabatic magnetization, state (3) reached by cooling in constant field and state (4) obtained after adiabatic demagnetization. See text for a more detailed discussion.}

\label{fig. 2}
\end{figure}

\begin{figure}
\includegraphics{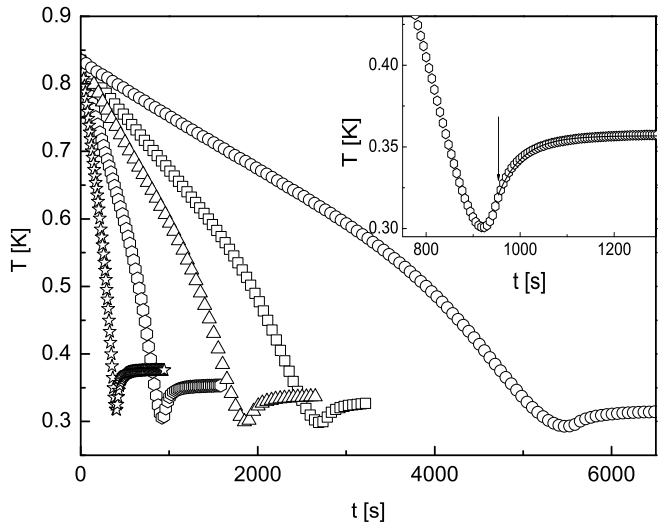} 
\caption{Time dependences of temperature during adiabatic demagnetization of Dy$_2$Ti$_2$O$_7$ for selected sweep rates:
(circles) - 7.8 mT/min, (squares) - 15.6 mT/min, (triangles) - 23.5 mT/min, (hexagons) - 47.1 mT/min, (stars)  - 110 mT/min. Inset: More detailed view of the time development of temperature at the end of the demagnetization. 
Solid line represents the least squares fit using Eq. (2). The moment when magnetic field reaches zero value is denoted by arrow.}
\label{fig. 3}
\end{figure}

\begin{figure}
\includegraphics{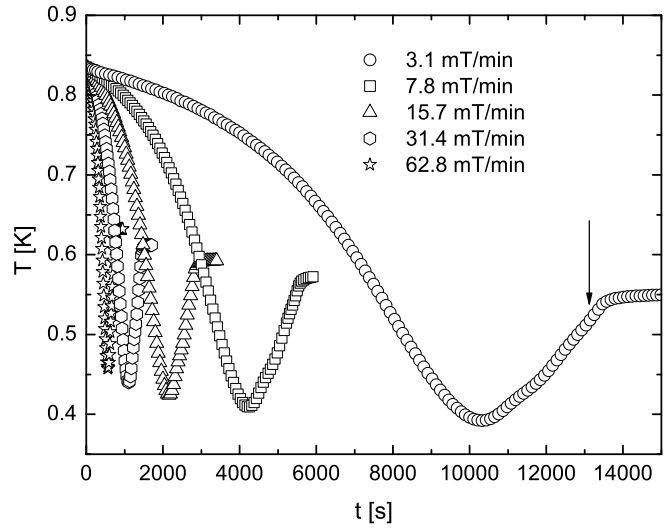} 
\caption{Time dependences of temperature during adiabatic demagnetization of (DyY)Ti$_2$O$_7$  for selected sweep rates. The arrow denotes the moment when magnetic field reaches zero value.}
\label{fig. 4}
\end{figure}

\begin{figure}
\includegraphics{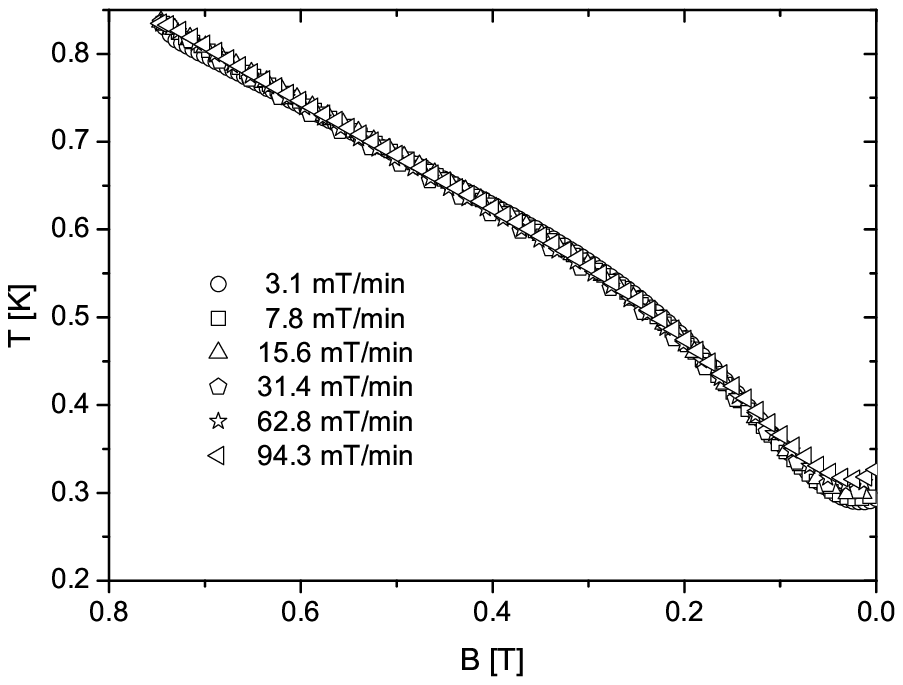} 
\includegraphics{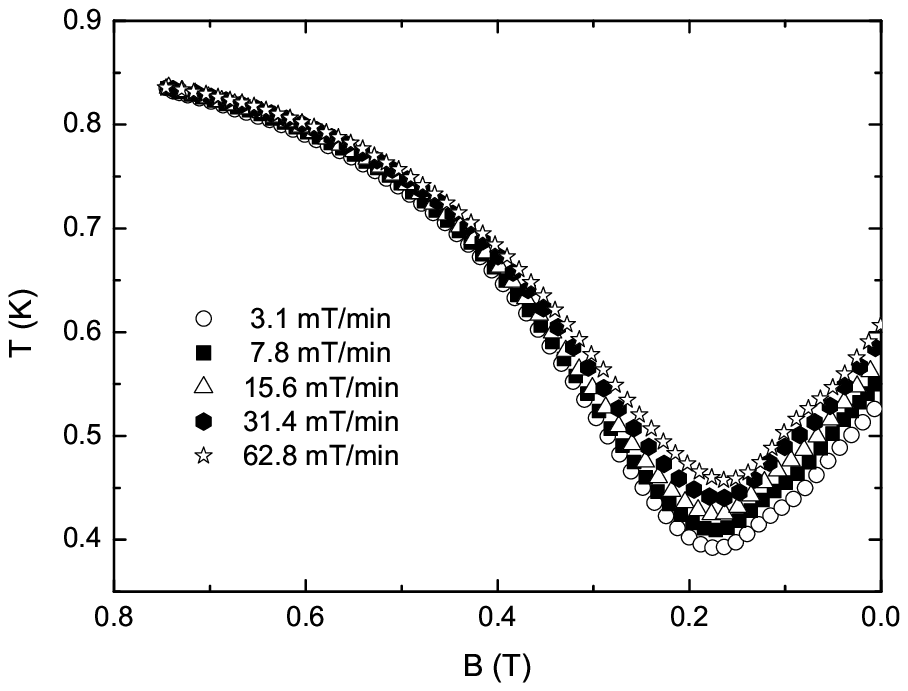} 
\caption{Magnetic field dependences of temperature during adiabatic demagnetization of 
a) Dy$_2$Ti$_2$O$_7$ and b) (DyY)Ti$_2$O$_7$  for corresponding sweep rates denoted in Fig. 3 and Fig. 4, respectively. The data of temperature are presented only to the point at which the magnetic field reaches zero value. See text for a more detailed discussion.}
\label{fig. 5}
\end{figure}

\begin{figure}
\includegraphics{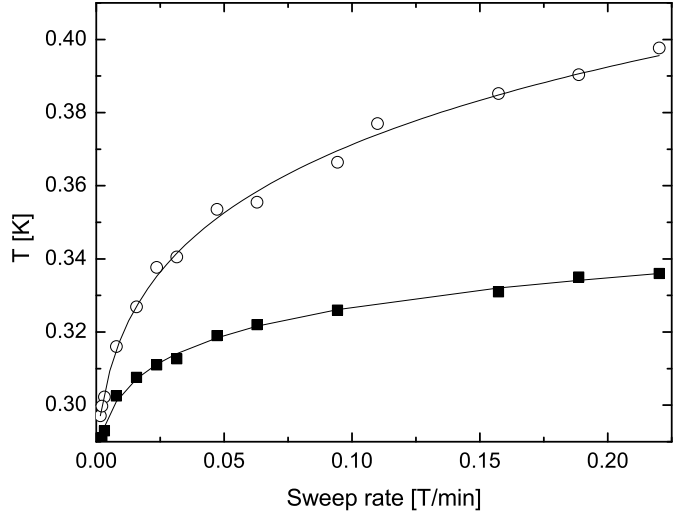} 
\caption{ Dependence of $T_{\rm final}$,  the resulting temperature after the demagnetization (empty circles) and $T_0$, the temperature at which magnetic field reached zero value (full squares), for Dy$_2$Ti$_2$O$_7$  on the used sweep rate. Three lowest temperatures were reached using 1.5 mT/min, 2.2 mT/min and 3.1 mT/min, respectively. 
The solid lines are guides for eye.}

\label{fig. 6}
\end{figure}

\begin{figure}
\includegraphics{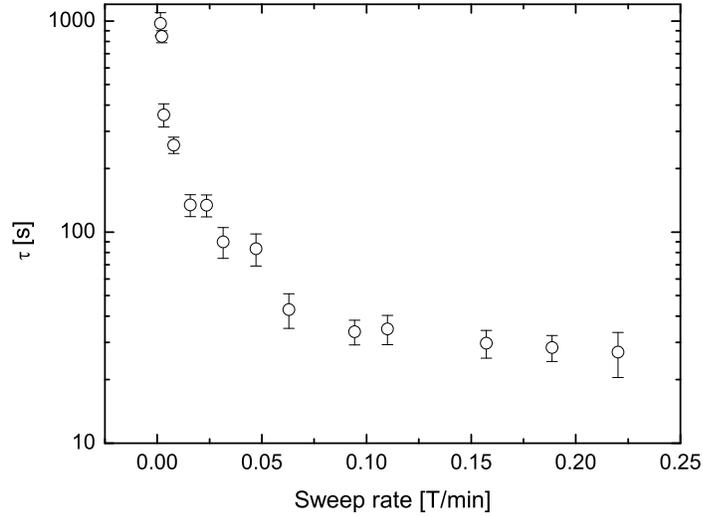} 
\caption{
Dependence of the estimated relaxation time of Dy$_2$Ti$_2$O$_7$ (empty circles) on the used sweep rate.}
\label{fig. 7}
\end{figure}

\begin{figure}
\includegraphics{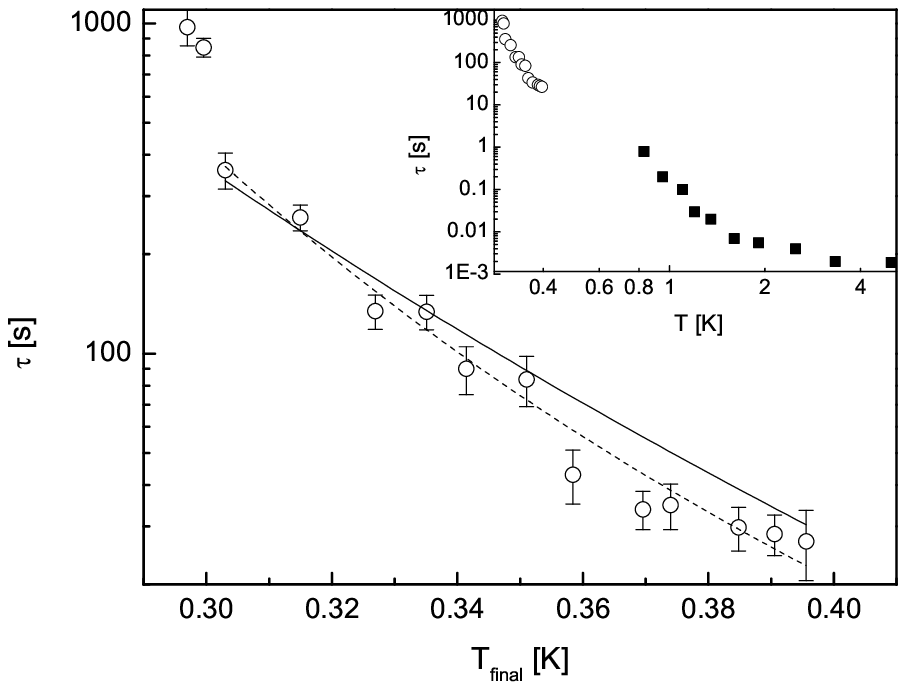} 
\caption{
Temperature dependence of the relaxation time of Dy$_2$Ti$_2$O$_7$ (empty circles). The solid line and dashed lines represent least squares fits from 0.3 K to 0.4 K using relations: $\tau=a.T^{-9}$, with $a=$ 7.22e-3 K$^9$s and $\tau=b*exp(\Delta/kT))$ with $b=$ 2.5e-3 s and $\Delta/k=$ 3.6 K, respectively.  
Inset: Comparison of the relaxation time obtained from the present work (empty circles) and from Ref. \onlinecite{Snyder1} (full squares).} 

\label{fig. 8}
\end{figure}


\begin{thebibliography} {5}

\bibitem{Ramirez} A. P. Ramirez, \textsl{Handbook of Magnetic Materials}, edited by K. J. H. Buschow (Elsevier, Amsterdam, 2001), Vol. 13; S. T. Bramwell and M. J. P. Gingras, Science {\bf 294}, 1495 (2001). 
\bibitem{Harris1} M. J. Harris, S. T. Bramwell, D. F. McMorrow, , T. Zeiske, and K. W. Godfrey, Phys. Rev. Lett., {\bf 79}, 2554 (1997).
\bibitem{Ramirez1} A. P. Ramirez, A. Hayashi, R. J. Cava, R. Siddharthan, and B. S. Shastry, Nature (London), {\bf 399}, 333, (1999). 
\bibitem{Kadowaki} H. Kadowaki, Y. Ishii, K. Matsuhira, and Y. Hinatsu, Phys. Rev. B {\bf 65}, 144421 (2002).  
\bibitem{Pauling} L. Pauling, J. Am. Chem. Soc. {\bf 57}, 2680 (1935). 
\bibitem{Fukazawa} H. Fukazawa, R. G. Melko, R. Higashinaka, Y. Maeno, and M. J. P. Gingras, Phys. Rev. B {\bf 65}, 054410 (2002).
\bibitem{Siddhartan1} R. Siddharthan, B. S. Shastry, A. P. Ramirez, A. Hayashi, R. J. Cava, and S. Rosenkranz, Phys. Rev. Lett. {\bf 83}, 1854 (1999).
\bibitem{Zhitomirsky} M. E. Zhitomirsky, A. Honecker, and O. A. Petrenko, Phys. Rev. Lett. {\bf 85}, 3269 (2000).
\bibitem{Moessner1} S. V. Isakov, R. Moessner, and S. L. Sondhi, Phys. Rev. Lett. {\bf 95}, 217201 (2005).
\bibitem{Ruff} J. P. C. Ruff, R. G. Melko, and M. J. P. Gingras, Phys. Rev. Lett. {\bf 95}, 097202 (2005).
\bibitem{Isakov} S. V. Isakov, R. Moessner, and S. L. Sondhi, Phys. Rev. Lett. {\bf 95}, 217201 (2005).
\bibitem{Sakakibara1} T. Sakakibara, T. Tayama, Z. Hiroi, K. Matsuhira, and S. Takagi, Phys. Rev. Lett. {\bf 90}, 207205 (2003).
\bibitem{Saito} M. Saito, R. Higashinaka, and Y. Maeno, Phys. Rev. B {\bf 72}, 144422 (2005).
\bibitem{Fennell} T. Fennell, O. A. Petrenko, B. Fak, J. S. Gardner, S. T. Bramwell, and B. Ouladdiaf, Phys. Rev. B {\bf72}, 224411 (2005).
\bibitem{Higashinaka} R. Higashinaka and Y. Maeno, Phys. Rev. Lett. {\bf95}, 237208 (2005).
\bibitem{Ehlers} G. Ehlers, A. L.Cornelius, M. Orend\'{a}\v{c}, M. Kaj\v{n}akov\'{a}, T. Fennel, S. T. Bramwell, and J. S. Gardner, J. Phys.: Condens. Matter {\bf15} L9 (2003).
\bibitem{Matsuhira2} K. Matsuhira, Y. Hinatsu, K. Tenya, and T. Sakakibara, J. Phys.: Condens. Matter {\bf12}, L649 (2000).
\bibitem{Matsuhira} K. Matsuhira, Y. Hinatsu, and T Sakakibara, J. Phys. Condens. Matter {\bf13}, L737 (2001).
\bibitem{Snyder2} J. Snyder, J. S. Slusky, R. J. Cava, and P. Schiffer, Nature {\bf413}, 48 (2001).
\bibitem{Snyder3} J. Snyder, J. S. Slusky, R. J. Cava, and P. Schiffer, Phys. Rev. B {\bf 66} 064432 (2002).
\bibitem{Snyder4} J. Snyder, B. G. Ueland, A. Mizel, J. S. Slusky, H. Karunadasa, R. J. Cava, and P. Schiffer, Phys. Rev. B {\bf 70}, 184431 (2004). 
\bibitem{Steve1} T. Fennell, O. A. Petrenko, B. Fak, S. T. Bramwell, M. Enjalran, T. Yavorskii, M. J. P. Gingras, R. G. Melko, and G. Balakrishnan, Phys. Rev. B {\bf 70}, 134408 (2004).
\bibitem{Steve2} T. Fennell, O. A. Petrenko, B. Fak, J. S. Gardner, S. T. Bramwell, and B. Ouladdiaf, Phys. Rev. B {\bf 72}, 224411 (2005).
\bibitem{Zhitomirsky1} M. E. Zhitomirsky, A. Honecker, and O. A. Petrenko, Phys. Rev. Lett. {\bf 85}, 3269(2000).
\bibitem{Zhitomirsky2} M. E. Zhitomirsky, Phys. Rev. B {\bf 67}, 104421 (2003).
\bibitem{Sosin} S. S. Sosin, L. A. Prozorova, A. I. Smirnov, A. I. Golov, I. B. Berkutov, O. A. Petrenko, G. Balakrishnan, and M. E. Zhitomirsky, Phys. Rev. B {\bf 71}, 094413 (2005).
\bibitem{Aoki} H. Aoki, T. Sakakibara, K. Matsuhira, and Z. Hiroi, J. Phys. Soc. Jpn. {\bf 73}, 2851 (2004).
\bibitem{Flood} We note one previous study of the adiabatic demagnetization of
Dy$_2$Ti$_2$O$_7$: D.~J. Flood, J. Appl. Phys, {\bf 45} 4041 (1974). 
\bibitem{from_Steve}R. S. Roth,  J. Res. Natl. Bur. Stand., {\bf 56}, 17 (1956). 
\bibitem{Mark} M. W. Meisel, G. R. Stewart, and E. D. Adams, Cryogenics {\bf 29}, 1168 (1989).
\bibitem{Uhlig} K. Uhlig, Cryogenics {\bf 35}, 525 (1995). 
\bibitem{Cryogenics} Ya. E. Volokitin, R. C. Thiel, and L. J.de Jongh, Cryogenic {\bf 34}, 771 (1994).
\bibitem{Snyder5} J. Snyder, B. G. Ueland, J. S. Slusky, H. Karunadasa, R. J. Cava, and P. Schiffer, Phys. Rev. B {\bf 69}, 064414 (2004). 
\bibitem{Snyder1} J. Snyder, B. G. Ueland, J. S. Slusky, H. Karunadasa, R. J. Cava, A. Mizel, and P. Schiffer, Phys. Rev. Lett. {\bf91}, 107201 (2003).
\bibitem{Matsuhira3} K. Matsuhira, Z. Hiroi, T. Tayama, S. Takagi, and T Sakakibara, J. Phys. Condens. Matter {\bf 14}, L559 (2002).
\bibitem{Orbach} R. Orbach, Proc. Roy. Soc. (London) {\bf A264}, 456 (1961).
\bibitem{ScottJeffries} P. L. Scott and C. D. Jeffries, Phys. Rev. {\bf 127}, 32 (1962). 
\bibitem{OrbachComment} The single ion Orbach process is almost certainly that which freezes out at 16 K~\cite{Snyder2}, identified as such by its Arrhenius activation energy close to the first set of excited crystal field states, $\Delta \approx 300 $ K. The large magnitude $\Delta$ makes this process unavailable at low temperature. 
\bibitem{Tom} T. Fennell, T. Yavorskii, S. T. Bramwell and M. J. P. Gingras, to be published.  
\bibitem{Melko} R. G. Melko, B. C. den Hertog, and M. J. P. Gingras, Phys. Rev. Lett. {\bf 87}, 067203 (2001).
\bibitem{Gardner} J. S. Gardner, A. Keren, G. Ehlers, C. Stock, E. Segal, J. M. Roper, B. Fak, M. B. Stone, P. R. Hammar, D. H. Reich, and B. D. Gaulin, Phys. Rev. B {\bf 68} 180401(R) (2003). 
\end{thebibliography}
\end{document}